\documentclass[12pt]{iopart}
\usepackage{graphicx}
\usepackage{amssymb}
\usepackage{color}
\usepackage{setstack}
\usepackage{iopams}
\usepackage{soul}
\def\Msol{{M_\odot}}
\newcommand{\drond}[2][ ]{\frac{\partial #1}{\partial #2}}

\newcommand{\deriv}[2][ ]{\frac{\mathrm{d} #1}{\mathrm{d} #2}}

\renewcommand{\epsilon}{\varepsilon}
\renewcommand{\phi}{\varphi}

\def\CC{{C\nolinebreak[4]\hspace{-.05em}\raisebox{.4ex}{\tiny\bf ++} }}
\def\R{{\mathbb R}}
\def\N{{\mathbb N}}
\newcommand\Nt{N_\theta}
\newcommand\Np{N_\phi}
\newcommand\Ylm{Y_{\ell m}}

\begin{document}

\title[]{ROXAS: a new pseudospectral non-linear code for general relativistic oscillations of fast rotating isolated neutron stars.}

\author{Ga\"el Servignat$^{1,3}$ and J\'er\^ome Novak$^{2,3}$}
\address{$^{1}$ Université Paris Cité, CNRS, Astroparticule et Cosmologie, F-75013 Paris, France\\
$^{2}$ Université de Strasbourg, CNRS, Observatoire astronomique de Strasbourg, UMR 7550, F-67000 Strasbourg, France\\
$^{3}$ Laboratoire Univers et Th\'eories, Observatoire de Paris, Universit\'e PSL, CNRS, Universit\'e Paris-Cit\'e, 92190 Meudon, France} 

\begin{abstract}
  Next-generation gravitational wave detectors are expected to increase their sensitivity in the kHz band where binary neutron star remnants are expected to emit. In this context,
  robust predictions of oscillation modes of the post-merger object are desirable. To that end, we present here our code \texttt{ROXAS} that is aimed at simulating isolated rotating
  neutron stars. It is based on our previously published formalism relying on primitive variables, along with the extended conformal-flatness condition formulation of Einstein equations.
  The equations are written under a well-balanced formulation.
  The code uses pseudospectral methods for both the metric and the hydrodynamics.
  Since standard test beds were already presented in a previous paper, we here mainly focus on presenting the numerical ingredients of \texttt{ROXAS}, as well as on the frequency extraction for 
  axisymmetric and non-axisymmetric modes. We compare our results with semi-analytic perturbative methods. Spherically symmetric modes are also recovered.
  The code currently supports fast, rigidly rotating isolated
  neutron stars, described with a cold equation of state. A notable feature of the code is its ability to follow the surface of the star at every step. It is very lightweight and can
  be run on office computers.
\end{abstract}

\submitto{\CQG}
\maketitle

\section{Introduction}
\label{sec:intro}
The LIGO-Virgo-Kagra collaboration has been observing compact binary mergers since almost a decade~\cite{abbott_gw150914_2016}. Three observation campaigns have led the collaboration to release a catalogue of
approximately 90 events~\cite{abbot_gwtc-3_2023}, and the fourth campaign is currently ongoing. Successors of current detectors, including Einstein Telescope~\cite{punturo_einstein_2010,maggiore_science_2020,
branchesi_science_2023} and Cosmic Explorer~\cite{reitze_cosmic_2019,evans_horizon_2021} will
have an increased sensitivity in the kHz band where the post-merger remnant of binary neutron stars (BNS) is expected to emit. Depending on the masses of the binary components, this object
is thought to be a metastable hypermassive neutron star (HMNS), supported by its rotation, therefore surviving for a few tenths of seconds before the rotation slows down and the HMNS
collapses to a black hole, see~\cite{metzger_kilonovae_2020} for a review. Before collapsing, the HMNS emits a large quantity of gravitational waves that, if detected, would help obtain
information about the interior of the neutron star (NS), and therefore on the equation of state (EoS). Having robust predictions of these oscillation modes is therefore a scientific goal
for the collaboration. To that end, we developed a new code, \texttt{ROXAS} (\textbf{R}elativistic \textbf{O}scillations of non-a\textbf{X}isymmetric neutron st\textbf{A}r\textbf{S}), dedicated to
the simulation of an isolated NS with the aim to provide oscillation mode frequencies of a HMNS. It is based for the hydrodynamics on the formalism developed in~\cite{servignat_new_2023}
that uses primitive variables (velocity and thermodynamic quantities) to write evolution equations of the enthalpy and velocity,
and on the extended conformal-flatness condition formulation (xCFC) formulation of Einstein equations developed in~\cite{cordero-carrion_improved_2009}. As most existing codes~\cite{cipolletta_spritz_2021,cipolletta_spritz_2022,
dimmelmeier_combining_2005,dimmelmeier_non-linear_2006,thierfelder_numerical_2011,pakmor_improving_2016,lioutas_general_2024,kidder_spectre_2017,rosswog_sphincs_bssn_2021,yamamoto_simulating_2008}
rely on conserved schemes~\cite{banyuls_numerical_1997}, a recovery step is needed to recover the primitive variables at each time step. Being very time-consuming,
and a possible source of code failure, configurations that do not exhibit shocks might rely on non-conserved formulations. The present code combines the advantages of spectral methods
and primitive variables to be as lightweight as possible. A spherically symmetric version of the code has been tried and tested in~\cite{servignat_new_2023},
so that we focus on the extension of the code to two and three spatial dimensions, and test it through the extraction of axisymmetric and non-axisymmetric oscillation mode frequencies.

The paper is organized as follows: in section~\ref{sec:equations}, we introduce the fluid variable notations and thermodynamic properties, write the evolution equations and then reformulate
them for numerical implementation. We also introduce the quadrupole formula and some of its extensions that account for general relativistic effects. In section~\ref{sec:numerical}, we describe the numerical
ingredients of \texttt{ROXAS}: three-dimensional grids, numerical variables, regularization techniques, boundary conditions, domain matching, interpolation techniques and the evolution algorithm; in
section~\ref{sec:tests} we show the results of the frequency extraction that we compare to previously published values. We summarize and conclude in section~\ref{sec:ccl}.

Unless otherwise stated, we use the set of geometrical units where $G=c=1$ and a 4-metric $g_{\mu\nu}$ of signature $(-,+,+,+)$, of which the associated covariant derivative (connection)
is denoted $\nabla_\mu$. We use Einstein's convention of summation over repeated indices. Greek indices are running from 0 to 3, whereas Latin lowercase ones are running from 1 to 3.
Latin uppercase indices are used to denote various thermodynamic species present in the fluid. Unit vectors in $\R^3$ are denoted with a hat, e.g. the orthonormal spherical triad is denoted
$(\hat{e}_r,\,\hat{e}_\theta,\,\hat{e}_\phi)=\left(\drond[]{r},\,\frac{1}{r}\drond[]{\theta},\,\frac{1}{r\sin\theta}\drond[]{\phi}\right)$.

\section{Theoretical background}
\label{sec:equations}
In this section we give the definitions and notations that we use to describe NS evolution. We give the thermodynamic assumptions, define the useful hydrodynamic quantities,
and write the 3+1 decomposition of spacetime that we use in \texttt{ROXAS}. Although mostly repeated from~\cite{servignat_new_2023}, the first three subsections are provided for completeness.
We then provide the set of equations for the hydrodynamic and metric sectors. Finally, the quadrupole formula, used to extract gravitational waveforms in the code, as well as some of its
extension taking General Relativity into account, are presented.

\subsection{Thermodynamics}

We consider a perfect fluid carried by the timelike unitary 4-velocity $u^\mu$ that carries all species and that satisfies $u^\mu u_\mu = -1$. The energy-momentum tensor of
a perfect fluid is:
\begin{equation}
  T^{\mu\nu} = (e+p)u^\mu u^\nu + pg^{\mu\nu},
\end{equation}
where $e$ is the energy density in the fluid frame and $p$ the pressure. The fluid is assumed to be composed of baryons at zero temperature and in $\beta$ equilibrium. 
Out-of-$\beta$-equilibrium stars can be described by adding appropriate evolution equations~\cite{servignat_new_2023,servignat_one-_2024}.
The thermodynamics of the fluid can therefore be described with a barotropic EoS, \textit{i.e.} that takes a single thermodynamic parameter as an argument.
The baryons, which have a mass $m_B$, are described by their number density $n_B$. We define the log-enthalpy $H$:
\begin{equation}
  H = \ln\left(\frac{e+p}{m_Bn_B}\right),
\end{equation}
and the sound speed $c_s$ which, in the barotropic case, is given by:
\begin{equation}
  c_s^2 = \deriv[p]{e}.
\end{equation}

\subsection{3+1 decomposition}

We consider an asymptotically flat spacetime. The spacetime metric is given in the standard 3+1 form:
\begin{equation}
	g_{\mu\nu}\mathrm{d}x^\mu\mathrm{d}x^\nu := -N^2\mathrm{d}t^2 + \gamma_{ij}(\mathrm{d}x^i + \beta^i\mathrm{d}t)(\mathrm{d}x^j + \beta^j\mathrm{d}t)
,\end{equation}
where $N$ is the lapse function, $\beta^i$ the shift 3-vector and $\gamma_{ij}$ the induced 3-metric on spacelike 3-hypersurfaces, associated with its covariant derivative $D_i$. The {extrinsic curvature} tensor is defined by
\begin{equation}
    K_{ij} := -\frac{1}{2}\mathcal{L}_\mathbf{n}\gamma_{ij}
,\end{equation}
where $\mathcal{L}_\mathbf{n}$ is the Lie derivative along the vector $n^\mu$, a unitary, future-oriented, timelike 4-vector normal to a hypersurface $\Sigma_t$ of constant time $t$:
\begin{equation}
    n_\mu := N\nabla_\mu{t}.
\end{equation}
$K:={K^i}_i$ denotes the trace of the extrinsic curvature tensor. Finally, we consider the following hydrodynamic quantities: the Lorentz factor with respect to the Eulerian observer will be denoted $\Gamma$, the Eulerian velocity $U^\mu$ and the coordinate velocity $v^i$. These quantities are related according to:
\begin{equation}
	\Gamma := -n_\mu u^\mu = (1-U_iU^i)^{-1/2}
,\end{equation}
and
\begin{equation}
	u^\alpha = \Gamma(n^\alpha + U^\alpha) = \frac{\Gamma}{N}(1,v^i),\quad U^i=\frac{1}{N}(v^i+\beta^i)
.\end{equation}

\subsection{Conformal decomposition and choice of coordinates}
In order to fix the coordinates, we must choose a gauge and a foliation. Following~\cite{bonazzola_constrained_2004}, we use the conformal decomposition of spacelike hypersurfaces. Let $f_{ij}$ be a flat metric,
$\bar{D}_i$ its associated covariant derivative, $\gamma = \det(\gamma_{ij})$ and $f = \det(f_{ij})$. We require that $\gamma_{ij}$ is $f_{ij}$ at spatial infinity. By introducing the so-called conformal factor
\begin{equation}
	\Psi = \left(\frac{\gamma}{f}\right)^{1/12}
,\end{equation}
the conformal metric is defined as
\begin{equation}
	\tilde{\gamma}_{ij} = \Psi^{-4}\gamma_{ij}
.\end{equation}
We choose the maximal slicing as a foliation:
\begin{equation}
	K = 0
,\end{equation}
and the {Dirac} gauge:
\begin{equation}
	\bar{D}_i\tilde{\gamma}^{ij} = 0
.\end{equation}

\subsection{Hydrodynamics sector}

The hydrodynamics is based on the formalism of~\cite{servignat_new_2023}, which uses only primitive variables. With all the hydrodynamic and thermodynamic quantities at hand,
the evolution equations of the log-enthalpy $H$ and the Eulerian velocity $U_i$ are:
\begin{eqnarray}\label{eq:evolH}
	\partial_tH = & -v^iD_iH - c_s^2\frac{\Gamma^2N}{\Gamma^2 - c_s^2(\Gamma^2-1)}\left[K_{ij}U^iU^j + D_iU^i - \frac{U^i}{\Gamma^2}D_iH\right] \\
  \label{eq:evolUi}
\partial_tU_i
   = & -v^jD_jU_i  + U_jD_i\beta^j - D_iN + U_iU^jD_jN \nonumber \\
	& + \frac{c_s^2\, N U_i}{\Gamma^2-c_s^2(\Gamma^2-1)}D_jU^j + U_i\frac{\Gamma^2(c_s^2-1)}{\Gamma^2-c_s^2(\Gamma^2-1)}NU^lU^jK_{lj} \nonumber\\
	& - \frac{N}{\Gamma^2}\left(D_iH - \frac{\Gamma^2(1-c_s^2)}{\Gamma^2-c_s^2(\Gamma^2-1)}U_iU^jD_jH\right)
.\end{eqnarray}
For a more robust numerical evolution, these equations will be rewritten in order to mathematically enforce equilibrium conditions. The first integral of motion of a general
relativistic rotating self-gravitating object is, for an axisymmetric stationary spacetime~\cite{bonazzola_axisymmetric_1993}:
\begin{equation}
  \label{eq:firstintegralofmotion}
  D_i\left(H_\mathrm{eq} + \ln{N_\mathrm{eq}} - \ln\Gamma_\mathrm{eq}\right) = 0.
\end{equation}
Here, the subscript $\mathrm{eq}$ indicates that it describes an equilibrium state. This expression is implicitly present in~(\ref{eq:evolUi}). It is possible to make it appear explicitly to remove the sum of gradients. In order to do so,
we write that the hydrodynamic, thermodynamic and metric quantities are the sum of their equilibrium value and the deviation to the equilibrium state (which is not necessarily small):
\begin{eqnarray}
  U_i     = & U_{i,\mathrm{eq}} + \bar{U}_i, \label{eq:expandUi}\\
  H       = & H_\mathrm{eq} + \bar{H}, \\
  \beta^i = & \beta^i_\mathrm{eq} + \bar{\beta}^i, \\
  N       = & N_\mathrm{eq} + \bar{N}. \label{eq:expandN}
\end{eqnarray}
Note that since the conformal factor $\Psi$ does not appear explicitly in the hydrodynamic equations, we do not need to write $\Psi=\Psi_\mathrm{eq}+\bar{\Psi}$.
In particular, the coordinate velocity $v^i$ is transformed into
\begin{equation}
  v^i = v^i_\mathrm{eq} + \bar{v}^i,
\end{equation}
where
\begin{eqnarray}
  v^i_\mathrm{eq} = & N_\mathrm{eq}U^i_\mathrm{eq} - \beta^i_\mathrm{eq} \\ 
  \bar{v}^i = & -\bar{\beta}^i + N_\mathrm{eq}\bar{U}^i + \bar{N}U^i_\mathrm{eq} + \bar{N}\bar{U}^i.
\end{eqnarray}
The equilibrium quantities are static: $\partial_tU_{i,\mathrm{eq}}=\partial_tH_\mathrm{eq}=0$. Two more equilibrium conditions can be enforced, based on the assumption of rigid rotation.
In this particular case, the coordinate velocity components are, in a spherical triad:
$v^r=v^\theta=0$, $v^\phi = \Omega r\sin\theta$, with $\Omega$ a rotation rate in $\mathrm{rad}\cdot\mathrm{s}^{-1}$ (if $f$ is the rotation frequency in $\mathrm{Hz}$, $\Omega=2\pi f$).
The conditions are:
\begin{eqnarray}
  \label{eq:nulldivergenceeq}
  D_jU^j_\mathrm{eq} = 0, \\
  \label{eq:relativisticvgradvplusgradv2}
  v^j_\mathrm{eq}D_jU_{i,\mathrm{eq}} + U_{j,\mathrm{eq}}D_iv^j_\mathrm{eq} = 0.
\end{eqnarray}
The condition~(\ref{eq:relativisticvgradvplusgradv2}) can be shown analytically when assuming that the conformal metric is flat; we will explain in the next subsection that we place the
present study within this assumption. Then, we write that $\beta^i = NU^i-v^i$, use the identity $D_i\Gamma = \Gamma^3U_jD_iU^j$ and inject the conditions~(\ref{eq:firstintegralofmotion}),~(\ref{eq:nulldivergenceeq})
and~(\ref{eq:relativisticvgradvplusgradv2}) to obtain the evolution equation of $\bar{U}_i$:
\begin{eqnarray}
  \label{eq:evolbarUi}
  \partial_t\bar{U}_i = & -\Big[v^j_\mathrm{eq}D_j\bar{U}_i + \bar{v}^jD_jU_{i,\mathrm{eq}} + \bar{v}^jD_j\bar{U}_i + U_{j,\mathrm{eq}}D_i\bar{v}^j + \bar{U}_jD_iv^j_\mathrm{eq} + \bar{U}_jD_i\bar{v}^j \nonumber\\
                        & - N\left(U_{j,\mathrm{eq}}D_i\bar{U}^j + \bar{U}_jD_iU^j_\mathrm{eq} + \bar{U}_jD_i\bar{U}^j\right)\Big] - \frac{N}{\Gamma^2}D_i\left(\bar{H} + \ln\left(1+\frac{\bar{N}}{N_\mathrm{eq}}\right)\right) \nonumber\\
  & + U_iU^jD_jN  + \frac{N}{\Gamma^2}\frac{\Gamma^2(1-c_s^2)}{\Gamma^2-c_s^2(\Gamma^2-1)}U_iU^jD_jH \nonumber\\
	& + \frac{c_s^2\, N U_i}{\Gamma^2-c_s^2(\Gamma^2-1)}D_j\bar{U}^j + U_i\frac{\Gamma^2(c_s^2-1)}{\Gamma^2-c_s^2(\Gamma^2-1)}NU^lU^jK_{lj}.
\end{eqnarray}
The evolution equation of $\bar{H}$ is:
\begin{eqnarray}
  \label{eq:evolbarH}
  \partial_t\bar{H} = & -\bar{v}^iD_iH_\mathrm{eq} - v^i_\mathrm{eq}D_i\bar{H} - \bar{v}^iD_i\bar{H} \nonumber\\
                      & - c_s^2\frac{\Gamma^2N}{\Gamma^2 - c_s^2(\Gamma^2-1)}\left[K_{ij}U^iU^j+ D_i\bar{U}^i - \frac{U^i}{\Gamma^2}D_iH\right].
\end{eqnarray}
This way to write the evolution equation is referred to in the literature as a \textit{well-balanced} formulation. Since we evolve only the "perturbation", this formulation helps
exploit all digits of the numerical representation of double precision floating-points, and allows as shown above to enforce equilibrium conditions exactly. Dumbser \textit{et al.}~\cite{dumbser_well-balanced_2023} 
provide a short review on well-balanced formulations. The evolution equations~(\ref{eq:evolbarUi}) and~(\ref{eq:evolbarH}) are the ones that are implemented in the code.

\subsection{Metric sector}

The metric quantities $N$, $\Psi$, $\beta^i$ and $\gamma_{ij}$ all enter (explicitly or implicitly) the hydrodynamics of a perfect fluid. Moreover, the spacetime responds to changes of its
matter content, meaning that evolving $H$ and $U_i$ should have an effect on the metric. The Einstein equations are therefore needed to compute the metric along the hydrodynamics.
We use the fully constrained formalism of~\cite{bonazzola_constrained_2004} and assume the conformal flatness condition (CFC), which amounts to set $\tilde{\gamma}^{ij}=f^{ij}$. We use the extended
formulation of~\cite{cordero-carrion_improved_2009}; the Einstein equations reduce to a hierarchical set of two linear vector and two non-linear scalar Poisson-like partial differential equations (PDE). As described
in~\cite{servignat_new_2023}, the code comprises two distinct numerical representations of the conformal factor $\Psi$. The first one, denoted $\Psi_\mathrm{ev}$, is obtained from the following evolution equation:
\begin{equation}
  \label{eq:evolpsiev}
  \partial_t\ln\Psi_\mathrm{ev} = \beta^i\bar{D}_i\ln\Psi_c + \frac{1}{6}\bar{D}_i\beta^i.
\end{equation}
$\Psi_c$ is the second representation of the conformal factor, and is obtained from the numerical solution of the Hamiltonian constraint of the Einstein equations.
Since the CFC kills all radiative degrees of freedom of the spacetime, the code does not naturally produce gravitational waves. Gravitational waves can still be extracted from the simulations
thanks to Einstein's quadrupole formula.

\subsection{Gravitational waves extraction}
To make the dimension of the formulas clearer, we bring back $c$ and $G$ in this subsection. The quadrupole formula, as derived by Einstein~\cite{einstein_uber_1918}, relates the gravitation quadrupole radial field $h^{TT}_{ij}$ to the variations of the quadrupole mass tensor $Q_{ij}$, and is given by:
\begin{equation}
  \label{eq:quadrupoleformula}
  h^{TT}_{ij}(\mathbf{x}, t) = \frac{2G}{c^4r}P_{ij}^{kl}(\mathbf{n})\ddot{Q}_{kl}\left(t-\frac{r}{c}\right),
\end{equation}
where $\mathbf{n}=\mathbf{x}/r$ is the direction of propagation, $P_{ij}^{kl}$ a projection operator on the direction of propagation, and the dots correspond
to the derivative with respect to time. The quadrupole formula is defined in a purely Newtonian context, and $\rho=m_Bn_B$ is the rest-mass density of the matter.
It has one flaw from the numerical point of view: it involves a second
derivative with respect to time. Numerical derivatives tend to be expensive to evaluate and to degrade the accuracy of the representation, especially time derivatives (in contrast with
spatial derivatives which, in this context, are computed with spectral methods). We follow~\cite{mueller_gravitational_1997} that uses the so-called \textit{stress formula} to replace time derivatives with
spatial derivatives:
\begin{equation}
  \label{eq:stressformula}
  \ddot{Q}_{ij} = \int_{\R^3}\rho(\mathbf{x},t)(2\dot{x}_i\dot{x}_j-x_i\partial_j\Phi-x_j\partial_i\Phi)\,\rmd^3x,
\end{equation}
where $\rho=m_Bn_B$ is the rest-mass density of the matter, $(x_i)_{i=1,2,3}$ a coordinate system (e.g. $x_1=x,\,x_2=y,\,x_3=z$ for the Cartesian coordinates), and $\Phi$ a Newtonian gravitational potential. We then use the plus/cross polarization decomposition of 
$h^{TT}_{ij}$(see~\cite{misner_gravitation_1973}):
\begin{equation}
  h^{TT}_{ij}(\mathbf{x},t) = \frac{1}{r}\left(A_+\mathbf{\hat{e}_+} + A_\times\mathbf{\hat{e}_\times}\right),
\end{equation}
where
\begin{eqnarray}
  \mathbf{\hat{e}_+} = & \mathbf{\hat{e}_\theta}\otimes\mathbf{\hat{e}_\theta} - \mathbf{\hat{e}_\phi}\otimes\mathbf{\hat{e}_\phi},\\
  \mathbf{\hat{e}_\times} = & \mathbf{\hat{e}_\theta}\otimes\mathbf{\hat{e}_\phi} + \mathbf{\hat{e}_\phi}\otimes\mathbf{\hat{e}_\theta},
\end{eqnarray}
with $\otimes$ the tensor product.
If the observer is in the equatorial plane with respect to the source, and denoting $\ddot{I}_{ij}=\frac{G}{c^4}\ddot{Q}_{ij}$, the expressions of $A_+$ and $A_\times$ are:
\begin{eqnarray}
  A_+^e = & \ddot{I}_{zz} - \ddot{I}_{yy},\\
  A_\times^e = & -2\ddot{I}_{yz}.
\end{eqnarray}
To compute $\ddot{I}_{ij}$ we employ some improvements. The first one, devised by~\cite{dimmelmeier_relativistic_2002-1} is a weight method to compute the gravitational potential $\Phi$ to include some
relativistic effects. Let $\Phi_1$ be the pure Newtonian potential, i.e. computed as a solution of
\begin{equation}
  \Delta{\Phi} = 4\pi G\rho.
\end{equation}
In the weak field limit, the gravitational potential can be approximated in {CFC} as:
\begin{equation}
  \Phi_2 = \frac{1}{2}\left(1-\Psi^4\right).
\end{equation}
We then define, for some $a>0$:
\begin{equation}
  \Phi_w = \frac{\Phi_1 + a\Phi_2}{1+a}
\end{equation}
the weighted Newtonian gravitational potential. In practice, we follow~\cite{dimmelmeier_relativistic_2002-1}, and we set $a=1/2$. We use $\Phi_w$ in equation~(\ref{eq:stressformula}). The second improvement is presented in~\cite{cerda-duran_cfc_2005}. It amounts
to use the Eulerian velocity $U_i$ as a relativistic equivalent of $\dot{x}_i$, and to redefine the density:
\begin{equation}
  \rho^* = \Psi^6\Gamma\rho,
\end{equation}
which amounts to take into account the local geometry of the spacetime instead of integrating as if it were flat. The final
formula is:
\begin{eqnarray}
  A_+^e      & = \frac{2G}{c^4}\int_{\R^3}\rho^*(\mathbf{x},t)\left(U_z^2-U_y^2 + y\partial_y\Phi_w - z\partial_z\Phi_w\right)\,\rmd^3x, \label{eq:formulaAplus} \\
  A_\times^e & =-\frac{2G}{c^4} \int_{\R^3}\rho^*(\mathbf{x},t)\left(2U_yU_z - z\partial_y\Phi_w - y\partial_z\Phi_w\right)\,\rmd^3x. \label{eq:formulaAcross}
\end{eqnarray}

\section{Numerical representation}
\label{sec:numerical}
The numerical implementation of \texttt{ROXAS} is based upon \texttt{LORENE}~\cite{gourgoulhon_lorene_2016}, a dedicated \CC library that comes with built-in three-dimensional Poisson solvers, based on Chebyshev pseudospectral methods~\cite{grandclement_multidomain_2001,gottlieb_numerical_1989,boyd_chebyshev_2001,grandclement_spectral_2009},
and using spherical coordinates. Pseudospectral methods are used to compute the spatial derivatives in the sources of the evolution and structure equations.
The ideas described in Section 4 of~\cite{servignat_new_2023} are mostly the same here, up to one major difference: the present three-dimensional code comprises two distinct grids. The first one, referred to as the \textit{metric grid}
is used to solve the Einstein equations, and the second one, referred to as the \textit{hydro grid}, is where the primitive variables are evolved. Since the hydro grid follows the star's surface that
is not spherically symmetric, and since the three-dimensional Poisson solvers rely on the sphericity of the numerical domains, the metric grid and the hydro grid need to be distinct. We here give details on each of those two grids.
\subsection{Metric grid}
The metric grid consists in a spherical nucleus, an arbitrary number of spherical shells and a compactified external domain (CED). If $N$ is the total number of domains, and $(r_i)_{i=1,\dots,N-1}$
a set of $N-1$ real numbers such that
\begin{equation}
  0<r_1<\dots<r_{N-1}<+\infty,
\end{equation}
then the nucleus spans $[0,r_1]\times\mathcal{S}$, the $i$-th shell spans $[r_i,r_{i+1}]\times\mathcal{S}$ and the CED spans $[r_{N-1},+\infty]\times\mathcal{S}$,
where $\mathcal{S}=\{(\theta,\,\phi),\,\theta\in[0,\pi],\,\phi\in[0,2\pi)\}$. The physical variables $(r,\,\theta,\,\phi)$ are mapped to the numerical coordinates $(\xi,\,\theta',\,\phi')$
such that we always have $\theta=\theta'$ and $\phi=\phi'$, while $\xi\in[0,1]$ in the nucleus and $\xi\in[-1,1]$ in the shells and the CED. The radial variable is changed according to
\begin{equation}
\left\{\begin{array}{lll}
  r & = r_1\xi, & \mbox{(nucleus)} \\
  r & = \frac{r_{i+1}-r_i}{2}\xi + \frac{r_{i+1}+r_i}{2}, & \mbox{($i$-th shell)} \\
  r & = \frac{2r_{N-1}}{1-\xi}. & \mbox{(CED)}
\end{array}\right.
\end{equation}
This changes the radial differential operator:
\begin{equation}
  \left\{\begin{array}{lll}
    \drond[]{r} & = \frac{1}{r_1}\drond[]{\xi}, & \mbox{(nucleus)} \\
    \drond[]{r} & = \frac{2}{r_{i+1}-r_i}\drond[]{\xi}, & \mbox{($i$-th shell)} \\
    \drond[]{r} & = \frac{(\xi-1)^2}{2r_{N-1}}\drond[]{\xi}. & \mbox{(CED)}
  \end{array}\right.
\end{equation}
The metric grid is used for several purposes. The main one is to solve for the metric. The metric quantities $\beta^i$, $N$, $\Psi_c$ and $\Psi_\mathrm{ev}$ are defined on the metric grid.
The source of equation~(\ref{eq:evolpsiev}) is computed on the metric grid so that $\Psi_\mathrm{ev}$ is evolved there. The metric grid also serves to store the equilibrium values of the
initial profiles $H_\mathrm{eq}$, $U_{i,\mathrm{eq}}$, $N_\mathrm{eq}$ and $\beta^i_\mathrm{eq}$. The angular directions are set by the number of grid points in the $\theta$ and $\phi$ directions, respectively
denoted $\Nt$ and $\Np$. We consider the spatial symmetry that leaves scalar fields unchanged under the transformation
\begin{equation}
  (\theta,\,\phi) \longrightarrow (\pi-\theta,\,\phi+\pi).
\end{equation}
Under this assumption, the angular variables only need to span $\theta\in[0,\pi/2]$ and $\phi\in[0,\pi)$, and the angular collocation points are chosen at
\begin{eqnarray}
  \label{eq:angularcollocationpoints}
  \forall \ell\in\{0,\dots,\Nt-1\},\,\theta_\ell = \frac{\ell}{\Nt-1}\frac{\pi}{2},\\
  \forall m\in\{0,\dots,\Np-1\},\,\phi_m=\frac{m}{\Np}\pi.
\end{eqnarray}
Since in this the angular variables span smaller intervals, fewer points are needed to achieve the same resolution as in the non-symmetric case. All domains share the same angular directions.

\subsection{Hydro grid}
\label{subsec:hydrogrid}

\begin{figure}
  \centering
  \includegraphics[width=.8\textwidth]{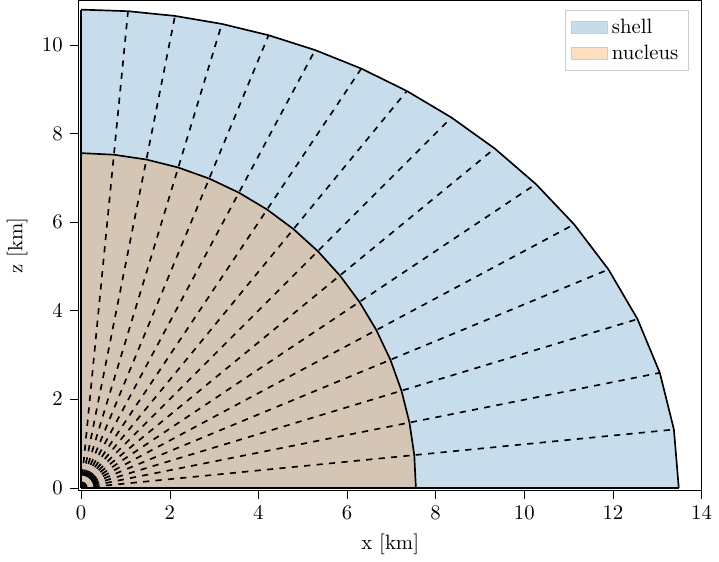}
  \caption[Rotating star wires]{Cross-section view of the rotating BU4 model (see Table~\ref{tab:seqBU}) in the $(x,z)$ plane. The nucleus
                                is spherical and the shell is adapted to the surface of the star. Each dashed line represents one of the interior angular directions of the grid.}
  \label{fig:starwires}
\end{figure}

The hydro grid consists in a spherical nucleus and an arbitrary number of shells. The last shell's outer boundary is deformed so that it is adapted to the surface of the star. Since most
simulations were run with exactly one shell we here write the mapping between the physical and numerical variables for two domains in the hydro grid. The case where there are no shells
is only used for spherically symmetric simulations and the code described in~\cite{servignat_new_2023} is to be preferred in this case. Let $r_\mathrm{nuc}$ be the radius of the nucleus,
and $R_S(\theta)$ the equilibrium coordinate radius profile of the rotating star. At equilibrium, the nucleus spans $[0,r_\mathrm{nuc}]\times\mathcal{S}$ and the shell spans
$[r_\mathrm{nuc},R_S(\theta)]\times\mathcal{S}$. This is illustrated in Figure~\ref{fig:starwires}. An important feature of the grid is that
the angular directions of the hydro grid are chosen to be the same as the ones of the metric grid. This helps increase the efficiency of interpolation procedures that we will describe in a further subsection. As the star evolves and oscillates, the 
outer boundary of the shell co-moves with the surface of the star. In order to do so, we impose that the value of $r$ on the outer boundary of the shell, denoted $R(t,\theta,\,\phi)$, is
evolved with the following PDE:
\begin{equation}
  \label{eq:impermeableBC}
  \drond[R]{t} = v_r(R(t,\,\theta,\,\phi),t) - \frac{v_\theta(R(t,\,\theta,\,\phi),t)}{R}\drond[R]{\theta} - \frac{v_\phi(R(t,\,\theta,\,\phi),t)}{R\sin\theta}\drond[R]{\phi}.
\end{equation}
The derivation of this equation is detailed in~\ref{app:derivationBC}, and the results of numerical experiments are shown in~\ref{app:numtestBC}: we perform simulations of rotating
ellipsoids and check for spectral convergence of the space scheme as well as convergence of the time integration scheme.

In the nucleus we map the physical coordinates $(t,\,r,\,\theta,\,\phi)$ to the numerical coordinates $(t',\,\xi,\,\theta',\,\phi')$ by:
\begin{equation}
  \left\{\begin{array}{rl}
    t & = t', \\
    r & = r_\mathrm{nuc}\xi, \\
    \theta & = \theta', \\
    \phi & = \phi'.
  \end{array}\right.
\end{equation}
The differential operators are then changed to:
\begin{equation}
  \left\{\begin{array}{rl}
    \drond[]{t} & = \drond[]{t'}, \\
    \drond[]{r} & = \frac{1}{r_\mathrm{nuc}}\drond[]{\xi}, \\
    \drond[]{\theta} & = \drond[]{\theta'}, \\
    \drond[]{\phi} & = \drond[]{\phi'}.
  \end{array}\right.
\end{equation}
In the shell we set:
\begin{equation}
  \left\{\begin{array}{rl}
    t & = t', \\
    r & = \frac{R(t',\,\theta',\,\phi')-r_\mathrm{nuc}}{2}\xi + \frac{R(t',\,\theta',\,\phi')+r_\mathrm{nuc}}{2}, \\
    \theta & = \theta', \\
    \phi & = \phi'.
  \end{array}\right.
\end{equation}
Then the differential operators become:
\begin{equation}
  \label{eq:diffopshelldeformed}
  \left\{\begin{array}{rl}
    \drond[]{t} & = \drond[]{t'} - \frac{2v_g}{R(t',\,\theta',\,\phi')-r_\mathrm{nuc}}\drond[]{\xi}, \\
    \drond[]{r} & = \frac{2}{R(t',\theta',\,\phi')-r_\mathrm{nuc}}\drond[]{\xi}, \\
    \drond[]{\theta} & = \drond[]{\theta'} - \frac{\xi+1}{R(t',\,\theta',\,\phi')-r_\mathrm{nuc}}\drond[R]{\theta'}\drond[]{\xi}, \\
    \drond[]{\phi} & = \drond[]{\phi'} - \frac{\xi+1}{R(t',\,\theta',\,\phi')-r_\mathrm{nuc}}\drond[R]{\phi'}\drond[]{\xi}.
  \end{array}\right.
\end{equation}
We have denoted with $v_g$ the \textit{grid advection velocity} that arises from using a moving grid (see~\cite{servignat_new_2023} for some more insight on grid advection). Its expression is
\begin{equation}
  v_g = \frac{1}{2}(\xi+1)\drond[R]{t'},
\end{equation}
and can therefore be computed thanks to the equation~(\ref{eq:impermeableBC}).

\subsection{Scalar fields regularity}
The simulations can exhibit a diverging behavior if two regularity conditions on scalar fields are not met. They are linked to the singularity of spherical coordinates on the origin and on
the vertical axis.
\subsubsection{Regularization at the origin}
The regularization scheme described in this paragraph was introduced by J. {Nicoules} in his
PhD thesis~\cite{nicoules_numerical_2023}. The spherical coordinates are degenerate at the origin, and therefore we expect that this particular physical point must receive a particular attention. An analytical field must not be
multivalued at the origin; numerically, the point $r=0$ is represented by as many grid point as there are angular directions. For example, if we choose $\Nt=5$ and $\Np=4$,
there are 20 grid points that represent the same $r=0$ physical point, and those grid points have \textit{a priori} no reason to remain identically valued during a simulation.
By changing how the division by $\xi$ is done in the nucleus, the divergences linked to multivalued fields at the origin can be cured. By default, the division by $r$ is done on the regular part of the field, i.e. if we want to divide a given function $f(r)$ by $r$, the
actual division that is done is $(f(r)-f(0))/r$. As in the nucleus, the radial variable $r$ is changed to the numerical variable $\xi$ by $r=r_\mathrm{nuc}\xi$, this corresponds to a division
by $\xi$, which is done in the coefficient space, thanks to the recurrence relation between the {Chebyshev} polynomials. To show the new procedure, we assume that $f$ is an even scalar field (if it is odd
$f(0)=0$ and there is no particular treatment):
\begin{equation}
	f(\xi) = \sum\limits_{k=0}^{N_r-1}a_iT_{2i}(\xi),
\end{equation}
with $T_{2i}$ the even Chebyshev polynomials and $N_r$ the number of collocation points in the radial direction.
Then, the expression $f(r)-f(0)$ can be interpreted in the coefficient space as $f(\xi) - f(0)T_0(\xi)$, which amounts to change the first coefficient of the spectral representation of $f$.
This operation can be performed on the last coefficient by writing $f(\xi)\leftarrow f(\xi)-f(0)T_{2N_r-1}(\xi)/T_{2N_r-1}(0)$, which looks on the coefficient like:
\begin{equation}\label{eq:regularization_divr}
	a_{N_r-1} \leftarrow a_{N_r-1} - (-1)^{N_r-1}\sum\limits_{k=0}^{N_r-1}(-1)^ka_k.
\end{equation}
This procedure has been implemented as the default one in \texttt{LORENE}.

\subsubsection{Regularization on the vertical axis}
\label{subsubsec:regularizationaxis}
Once again, analytical fields must not be multivalued on the vertical axis; a grid point on the vertical axis has $\Np$ representatives. If a regular scalar field $f$ is expanded on
the spherical harmonics basis like
\begin{equation}
  \label{eq:expansionylm}
  f(r,\,\theta,\,\phi) = \sum\limits_{\ell=0}^{\ell_\mathrm{max}}\sum\limits_{m=-\ell}^\ell f_{\ell m}(r)\Ylm(\theta,\,\phi),
\end{equation}
then for $m\neq0$, $f_{\ell m}\Ylm$ must behave as $\sin^m(\theta)$ when $\theta\rightarrow0$. Because $f_{\ell m}$ only depends upon the radial variable and $\Ylm\propto\sin^m(\theta)$,
the condition is automatically fulfilled when the fields are represented on the spherical harmonics basis. Since derivation routines in \texttt{LORENE} are written on the Fourier basis of
sines and cosines, the physical fields are represented on this basis by default. This means that the regularization procedure consists in simply performing a basis change on scalar fields
from Fourier to spherical harmonics, and back to Fourier again. The transformation costs $\mathcal{O}(N\log{N})$ in each direction so if $N=\max(N_r,\Nt,\Np)$, this amounts to a procedure
that costs $\mathcal{O}((N\log{N})^3)$.

\subsection{Filters}
\label{subsec:filters}
As we have seen in Section~\ref{subsec:hydrogrid}, the angular differential operators involve an additional Jacobian term which depends on the angular variables. Since it is non-linear in $R$,
it may provoke aliasing issues as the multiplication is performed in the collocation space. To cure the issues associated with the Jacobian, we introduce exponential filters~\cite{canuto_spectral_2006,hesthaven_spectral_2007} that we apply on
the spectral representation $(f_{i\ell m})$ of a scalar field $f$ expanded on Chebyshev polynomials and spherical harmonics (which is simply considering in equation~(\ref{eq:expansionylm}) 
the expansion of $f_{\ell m}$ on Chebyshev polynomials). Given three positive integers $(p_r,p_\theta,\,p_\phi)$, we apply the following operation:
\begin{equation}
  \mathcal{F}(f_{i\ell m}) = f_{i\ell m}e^{-\alpha[i/(N_r-1)]^{p_r}}e^{-\alpha(\ell/\ell_\mathrm{max})^{p_\theta}}e^{-\alpha(m/\Np)^{p_\phi}},
\end{equation}
where $\alpha=\ln(10^{-16})$ so that the last coefficient is always set below the machine accuracy. It is possible to choose not to filter in a given direction; typically the enthalpy and velocity profiles are not filtered in the radial direction. The radial filter is used only on the sources
of the Einstein equations. This procedure can be directly applied on scalar fields such as the enthalpy profile, but to filter vector fields that are by default represented with the spherical
triad, we must first rotate the triad to get the Cartesian components of the vector field, filter those components and rotate the triad back to spherical.

\subsection{Boundary conditions}
\label{subsec:boundaryconditions}
\subsubsection{Angular boundary conditions}
Since the angular variables $(\theta,\,\phi)$ span $[0,\pi/2]\times[0,\pi)$, the numerical domain has borders where $\theta=0,\,\pi/2$ and $\phi=0,\pi$. These numerical borders are of
course not physical border, and the angular boundary conditions (BC) are taken care of with the Galerkin method that consists in choosing a basis that fulfills the desired BC so that
it is automatically fulfilled by the numerical representation of the field.

\subsubsection{External boundary conditions}
Here we describe the boundary conditions imposed on the outermost boundaries of the numerical domains.
On the metric grid, we impose the conditions that $\Psi\rightarrow1$, $N\rightarrow1$ and $\beta^i\rightarrow0$ for $r\to+\infty$ (corresponding to $\xi\to1$ in the CED) as the spacetime is asymptotically flat. On the hydro grid, we impose the
impermeable boundary condition that requires that no matter crosses the numerical domain outermost boundary. It is already accounted for by choosing the comoving grid with equation~(\ref{eq:impermeableBC})
(see also~\ref{app:derivationBC}).

\subsubsection{Domain matching}
Now the communication between the domains remains to be addressed. On the metric grid, the metric fields $\Psi_c$, $N$ and $\beta^i$ are matched directly by the Poisson solvers which require
the continuity of the functions and their first radial derivative. Equation~(\ref{eq:evolpsiev}) is an advection equation with characteristic speed $-\beta^r$ and is matched
in an upwind fashion: given two neighboring domains $l_z$ and $l_z+1$ and an angular direction, if $\beta^r<0$ the value of the last grid point of the domain $l_z$ is copied
to the first grid point of the domain $l_z+1$, and \textit{vice-versa} if $\beta^r>0$. On the hydro grid we match the enthalpy profile with the tau method. The deformed domain along with Equation~(\ref{eq:diffopshelldeformed})
implies that the angular coefficients of $R(\theta,\,\phi)$ are mixed with those of $\partial_\xi H$ so the spectral representation of $H$ cannot be used directly for this purpose.
Instead, we perform the inverse Fourier transform on the angular coefficients of $H$ so that the resulting object is a collection of radial functions $\tilde{H}_{\ell m}(r)=H(r,\,\theta_\ell,\,\phi_m)$,
where $\theta_\ell$ and $\phi_m$ are defined in Equation~(\ref{eq:angularcollocationpoints}). Then, every $\tilde{H}_{\ell m}$ is matched according to the procedure described in \ref{app:matchwires}.
Assuming a subsonic flow, there is one ingoing characteristic at the outer boundary of the nucleus and one at the inner boundary of the shell. We therefore impose the continuity of all
$\tilde{H}_{\ell m}$ and $\partial_r\tilde{H}_{\ell m}$.

Even though it is not motivated mathematically, we found that the equilibrium term $\bar{H}+\ln(1+\bar{N}/N_\mathrm{eq})$ (see Equation~(\ref{eq:evolUi})) also needed to be matched when
computing the sources of the hydrodynamic equations. We employ the procedure of~\ref{app:matchwires} again.

\subsection{Evolution algorithm}
We presented all the technical components of the \texttt{ROXAS} code and now turn to sum them up in a description of the successive steps to perform the simulation. Since there are two
distinct grids for the metric and the hydrodynamics, a number of interpolation steps are involved to compute the sources of the Einstein equations and of the hydrodynamic equations, that
we detail hereafter.
\begin{itemize}
  \item Computation of the source terms of the metric equations (see Equations~(30)-(33) of~\cite{cordero-carrion_improved_2009}).
  \begin{enumerate}
    \item We first interpolate $\Psi_\mathrm{ev}$ to the hydro grid to define $\gamma_{ij}=\Psi_\mathrm{ev}^4f_{ij}$ there, which allows to compute the {Lorentz} factor
    $\Gamma=(1-U_iU^i)^{-1/2}=(1-\gamma_{ij}U^iU^j)^{-1/2}$, and then to compute the matter terms $E=\Gamma^2(e+p)-p$, $S=3p+(\Gamma^2-1)(e+p)$ and $p_j=\Gamma^2(e+p)U_j$.
    \item Then we interpolate $E$ and $p_j$ to the metric grid and form the quantities $E^*=E\Psi_\mathrm{ev}^6$ and $p_j^*=p_j\Psi_\mathrm{ev}^6$.
    \item The first two metric equations can be solved. $\Psi_c$ is now known, and we compute $S^*=S\Psi_c^6$ where $S$ has been computed on the hydro grid and then 
          interpolated to the metric grid. The last two metric equations can be solved.
  \end{enumerate}
  \item Computation of the source terms of the hydrodynamic equations.
  \begin{enumerate}
    \item We interpolate $\Psi_c$ to the hydro grid to define $\gamma_{ij}=\Psi_c^4f_{ij}$.
    \item We interpolate the lapse and the shift on the hydro grid, but this time we separate the equilibrium and the bar quantities. $N_\mathrm{eq}$ and $\beta^i_\mathrm{eq}$
          are simply interpolated from the metric to the hydro grid, while to compute $\bar{N}$ and $\bar{\beta}^i$ we interpolate $N-N_\mathrm{eq}$ and $\beta^i-\beta^i_\mathrm{eq}$:
          these differences are computed on the metric grid before interpolation.
          The total lapse and shift on the hydro grid are formed with $N_\mathrm{eq}+\bar{N}$ and $\beta^i_\mathrm{eq}+\bar{\beta}^i$ where we use the interpolated values previously
          computed.
  \end{enumerate}
\end{itemize}

Then, the algorithm to perform a time step is the following:
\begin{enumerate}
  \item At a given time $t$, $H$, $U_i$ and $\Psi_\mathrm{ev}$ are known thanks to their evolution equations~(\ref{eq:evolUi}),~(\ref{eq:evolH}) and~(\ref{eq:evolpsiev}). $\Gamma$ can be computed from $U_i$, and along with the EoS,
  $H$ allows computing $p$ and $e$.
  \item Following the interpolation procedure we compute the sources of the Einstein equations and solve them with the relaxation iterative method. 
  The output is the value of $\Psi_c$ consistent with the Hamiltonian constraint, which replaces the value of $\Psi_\mathrm{ev}$ computed with Equation~(\ref{eq:evolpsiev})
  as soon as it is available, as well as the lapse $N$ and the radial component of the shift vector $\beta^r$.
  A number of simulations will be run using the Cowling approximation which amounts to fix the spacetime to its initial equilibrium value; in this case the current step is skipped.
  \item We use the knowledge of the radius of the star $R$, as well as $H,\,U_i,\,N,\beta^i$ and $\Psi_c$ together with the EoS and the interpolation procedure to compute the sources of Equations~(\ref{eq:evolUi}), (\ref{eq:evolH}), (\ref{eq:impermeableBC}) and~(\ref{eq:evolpsiev}).
  \item $H$, $U_i$, $R$ and $\Psi_\mathrm{ev}$ are evolved with an explicit time integration scheme.
  \item The hydro grid of the next time step is computed.
  \item The boundary conditions are imposed as described in Section~\ref{subsec:boundaryconditions}.
\end{enumerate}
The evolution is performed with an Adams-Bashforth scheme of order 3, which uses the knowledge of the current and previous two time steps to compute the next one. We therefore
initialize the first two time steps with a Runge-Kutta scheme of order 4 (RK4 scheme).
All initial data are computed using the initial data solvers of \texttt{LORENE}~\cite{gourgoulhon_fast_1999,lin_rotating_2006}. To trigger the oscillations, the star is perturbed with the profile given by:
\begin{equation}
  \label{eq:nonaxisymmetricperturbationdensity}
  \delta{H} = \epsilon(x^2\pm y^2)\left(1-\frac{r^2}{R_S(\theta)^2}\right).
\end{equation}
It is a mixture between $\ell=m=0$ and $\ell=2,\,m=0$ modes (+ sign) or $\ell=m=0$ and $\ell=m=2$ modes (- sign). The radial dependence of the perturbation profile is arbitrary, but we
choose it to be a low order polynomial so that it is easily represented by the spectral decomposition, as well as fulfilling the initial condition $H=0$ on the border of the star.
The parameter $\epsilon$ allows to control the strength of the perturbation. 

\begin{figure*}
  \centering
  \includegraphics[width=\textwidth]{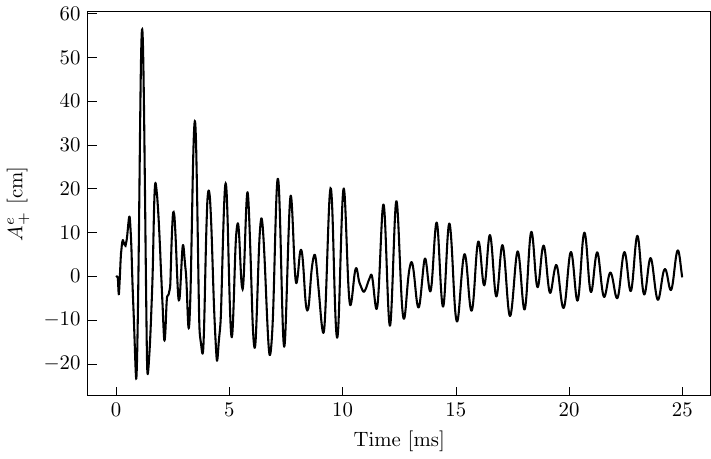}
  \caption[$A_+^e$ waveform of BU4 model]{$A_+^e$ {GW} waveform for the BU4 model. The {Fourier} spectrum is shown on Fig.~\ref{fig:spectrumBU4}. The value of $A_+^e$ corresponds,
          for a source at $10\,\mathrm{Mpc}$, to a strain $h_+\sim10^{-24}$.}
  \label{fig:waveformBU4}
\end{figure*}

\begin{figure*}
  \centering
  \includegraphics[width=\textwidth]{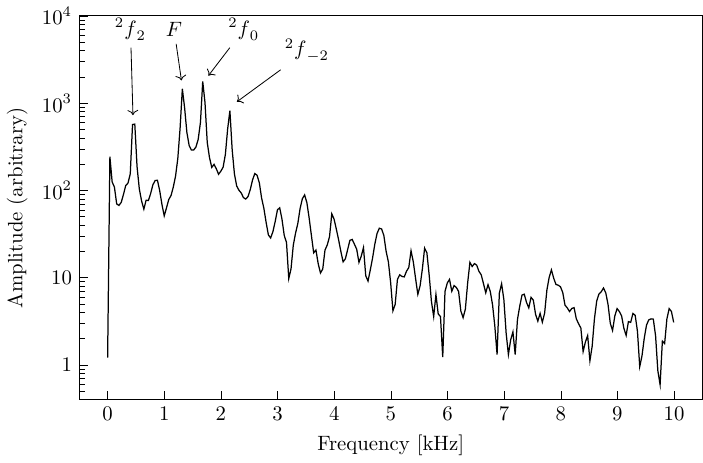}
  \caption[$A_+^e$ waveform spectrum of BU4 model]{{Fourier} spectrum of the signal on Fig.~\ref{fig:waveformBU4}, with the mode peaks highlighted.}
  \label{fig:spectrumBU4}
\end{figure*}

\section{Tests}
\label{sec:tests}
To study oscillation modes of rotating stars, we use the widely studied sequence of uniformly rotating polytropic stars BU~\cite{dimmelmeier_non-linear_2006,stergioulas_non-linear_2004,gaertig_oscillations_2008,kruger_oscillations_2010,zink_frequency_2010,kruger_dynamics_2020}.
In particular, Ref.~\cite{dimmelmeier_non-linear_2006} gives values for the frequencies of axisymmetric modes in {CFC}, Refs.~\cite{gaertig_oscillations_2008,kruger_oscillations_2010} give values for
the frequencies of axisymmetric and non-axisymmetric modes using the {Cowling} approximation and Ref.~\cite{kruger_dynamics_2020} gives values for the frequencies of non-axisymmetric modes in full {GR}. We give the macroscopic properties of the sequence BU in Table~\ref{tab:seqBU}.
Even though the mode frequencies are present in the waveforms computed with Eqs.~(\ref{eq:formulaAplus}) and~(\ref{eq:formulaAcross}),
we extract the frequencies from the appropriate coefficient of the decomposition of the radius on spherical harmonics, i.e. quasi-radial modes are extracted from the $\ell=m=0$ coefficient, 
axisymmetric modes from the $\ell=2,\,m=0$ coefficient and non-axisymmetric modes from the $\ell=m=2$ coefficient. The grid settings are the following:
\begin{itemize}
  \item The hydro grid is composed of two domains: a spherical nucleus which is fixed and set to span $[0,0.7r_p]$ where $r_p$ is the coordinate radius at $\theta=0$ i.e. the North Pole,
        and a shell that describes the rest of the star. We set $N_r=17$ in both domains, and $\Nt$ is set as stated in Table~\ref{tab:seqBU} (except for BU0 where we set $\Nt=9$). The 2D simulations are run with $\Np=1$ and
        all 3D simulations are run with $\Np=8$.
  \item In the metric grid, we set a nucleus and a shell that respectively span $[0,8\,\mathrm{km}]$ and 
        $[8\,\mathrm{km},r_p]$. Next we choose an arbitrary number of shells so that they span intervals of the form $[R_S(\theta_i),R_S(\theta_j)]$, $i<j$. Beyond $r_e$ we set a final shell and the CED. In the first two domains we set
        $N_r=25$; in the shells spanning $[r_p,r_e]$ we set $N_r=25$ or $33$ depending on the size of the domain; we finally set $N_r=25$ in the last shell and $N_r=17$ in the CED. The angular directions
        of the metric grid are the same as the hydro grid and therefore the values of $\Nt$ and $\Np$ are identical.
\end{itemize}
We once again employ exponential angular filters in the hydro grid. The value of the order for the $\theta$ filter is decreased (i.e. its strength is increased) as $\Nt$ increases; we use an order 10 filter for $\Nt=9$ and $17$, an order
8 filter for $\Nt=25$ and an order 6 filter for $\Nt=33$. When $\Np>1$, we always set $\Np=8$ and in that case we use an order 12 filter. We filter the enthalpy and Eulerian velocity profiles, as
well as the sources of the Einstein equations, as described in Sec.~\ref{subsec:filters}. For the non-rotating star BU0 we set only angular filters of order 20 and 22 for $\theta$ and $\phi$ respectively. For the rotating stars we set an order 4 radial filter,
an order 6 $\theta$ filter and an order 16 $\phi$ filter. The simulations are performed with a time step $\Delta{t} = 1.67\times10^{-7}\,[\mathrm{s}]$, except for the simulations of the BU7 model where we use $\Delta{t}=3.34\times10^{-8}\,[\mathrm{s}]$.
We give the extracted frequencies of the modes in Tables~\ref{tab:BUfreqsCFC} and~\ref{tab:BUfreqsCowling},
that we can compare to tabulated frequencies of the literature, reported in Table~\ref{tab:BUfreqsliterature}. We compute the relative differences between our computed values and the tabulated ones and report them in Table~\ref{tab:BUfreqsreldiff}.
Some relative differences are of the order of 3-4\% and some surpass 10\%, but the latter corresponds to some of the lowest frequencies ($\sim100-200\,\mathrm{Hz}$) for which we do not expect an accurate estimation with $25\,\mathrm{ms}$ simulations. Otherwise, most frequencies
are recovered at the level of 1\% or less. We therefore consider that the frequency extraction is successful. This success also justifies the {CFC} approach to extract oscillation frequencies, compared to full
{GR} evolutions that are typically much more computationally expensive. The lower frequencies could be re-extracted with longer simulation using future improved versions of \texttt{ROXAS}. We also note that
the frequencies of the $F$ and $H_1$ mode for the non-rotating model BU0 produced with the three-dimensional code (although extracted from two-dimensional simulations) are consistent with the frequencies
produced with the early spherically symmetric version of the code~\cite{servignat_new_2023}. We plot on Figures~\ref{fig:waveformBU4} and~\ref{fig:spectrumBU4} the {GW} signal from the simulation performed with the BU4 model, along with its {Fourier} spectrum, and highlight
the peaks with their corresponding mode.

Furthermore, we performed the simulation of a rotating star described with the barotropic SLy4 {EoS}~\cite{gulminelli_unified_2015}. To do so, we used the pseudo-polytropic fit described in~\cite{servignat_one-_2024} and for which the coefficients therein.
We choose a rotation frequency of $f=191\,\mathrm{Hz}$ with almost the same central enthalpy as the BU sequence: $H_c=0.22656$, so that it has a baryon mass of $M_b=1.504\Msol$,
a gravitational mass of $M=1.361\Msol$ and $\Omega/\Omega_K\approx16\%$. Since its equilibrium coordinate polar radius is $r_p=9.34\,\mathrm{km}$, we set the nucleus of the metric grid up to $7\,\mathrm{km}$ instead of $8\,\mathrm{km}$,
and set $N_r=33$ in all metric domains (except the CED where we keep $N_r=17$). The angular filter for the hydro grid are set at order 10 for the $\theta$ direction and 12 for the $\phi$ direction, and the filters
for the metric grid are set to order 4 (radial), order 6 ($\theta$) and order 16 ($\phi$). We use Eq.~(\ref{eq:nonaxisymmetricperturbationdensity}) with $\epsilon=10^{-6}$ to perturb the star, and
let it evolve for $25\,\mathrm{ms}$. We extract the frequencies of the ${^2f}_2$ and ${^2f}_{-2}$ modes for which we find frequencies at
$1.675\,\mathrm{kHz}$ and $2.185\,\mathrm{kHz}$. We compare with~\cite{kruger_dynamics_2020} where they study the same star, and they find $1.655\,\mathrm{kHz}$ and $2.105\,\mathrm{kHz}$ respectively,
which constitutes a relative difference of 1.2\% and 3.7\%. Once again the frequency extraction is satisfactory and this is an important step for the validation of the \texttt{ROXAS} code with realistic {EoS}s.
We also provide the frequencies of the other modes: the $F$ and ${^2f}_0$ are respectively found at frequencies of $3.085\,\mathrm{kHz}$ and $1.964\,\mathrm{kHz}$. We finally provide the values of the modes with simulations performed using
the {Cowling} approximation: the frequency extraction gives $4.415\,\mathrm{kHz}$, $2.400\,\mathrm{kHz}$, $2.141\,\mathrm{kHz}$ and $2.640\,\mathrm{kHz}$ for the $F$, ${^2f}_0$, ${^2f}_2$ and ${^2f}_{-2}$ modes respectively.

\begin{table*}
  \caption[BU rotating stars sequence]{Properties of the BU sequence as computed with the initial data solvers of \texttt{LORENE}~\cite{gourgoulhon_fast_1999,lin_rotating_2006}. All stars are built with a central enthalpy of $H_c = 0.227823853$.
                                       $\Nt$ is the number of points in the $\theta$ direction needed to describe the surface of the star. We give the gravitational mass $M$ and the circumferential equatorial radius $R$,
                                       as well as the ratio between the coordinate radii at the North Pole
                                       $r_p$ and at the equator $r_e$. $\Omega_K=2\pi\times853\,\mathrm{rad}\cdot\mathrm{s}^{-1}$  is the {Kepler} frequency of the star. The ratio $\Omega/\Omega_K$ is given
                                       in \%.}
  \label{tab:seqBU}
  \centering
  {\begin{tabular}{cccccccc}
    \hline
    Model & $f\,[\mathrm{Hz}]$ & $\Omega/\Omega_K$ & $M \,[\Msol]$ & $R\,[\mathrm{km}]$ & $r_e\,[\mathrm{km}]$ & $r_p/r_e$ & $\Nt$ \\
    \hline
    \hline
    BU0 & 0   & 0 & 1.400 & 14.15 & 11.99 & 1 & 1 \\
    \hline
    BU1 & 347 & 41 & 1.431 & 14.51 & 12.30 & 0.950 & 9  \\
    \hline
    BU2 & 487 & 57 & 1.465 & 14.92 & 12.64 & 0.900 & 9 \\
    \hline
    BU3 & 590 & 69 & 1.502 & 15.37 & 13.03 & 0.850 & 17 \\
    \hline
    BU4 & 673 & 79 & 1.542 & 15.90 & 13.49 & 0.800 & 17 \\
    \hline
    BU5 & 740 & 87 & 1.585 & 16.51 & 14.03 & 0.750 & 25 \\
    \hline
    BU6 & 793 & 93 & 1.627 & 17.26 & 14.70 & 0.700 & 25 \\
    \hline
    BU7 & 831 & 97 & 1.665 & 18.16 & 15.55 & 0.649 & 33 \\
    \hline
    \hline
  \end{tabular}}
\end{table*}

\begin{table}
  \caption[Oscillation frequencies (BU sequence, {CFC})]{Oscillation frequencies of the BU sequences, solving for the metric. The frequencies are given in $\mathrm{kHz}$. The $F$ mode is the fundamental quasi-radial mode,
  the $H_1$ mode is its first overtone, the ${^2f}_0$ mode is the fundamental $\ell=2$, $m=0$ mode and the ${^2f}_2$ and ${^2f}_{-2}$ modes are the counter- and corotating $\ell=m=2$ modes. All frequencies are obtained with the 
  perturbation~(\ref{eq:nonaxisymmetricperturbationdensity}) using $\epsilon=10^{-2}$. The displayed frequencies for the $F$, $H_1$ and $^2f_0$ mode are obtained with 2D simulations,
  but we checked that the 3D simulations gave consistent frequencies for these particular modes.
  $\delta{f}_{2D,3D}=1/t_\mathrm{max}$ are the discretization steps in the {Fourier} space, for 2D and 3D simulations.
  It is dictated by the length of the simulation, $t_\mathrm{max}$ being the final simulation time. The ${^2f}_2$ mode of the BU7 star has not been extracted as its expected frequency is
  approximately 11 Hz i.e. a period of 91 ms, which the simulations of 25 ms do not resolve.}
  \label{tab:BUfreqsCFC}
  \centering
  {\begin{tabular}{cccccccc}
    \hline
    Model & $F$ & $H_1$ & ${^2f}_0$ & $\delta{f}_{2D}\,[\mathrm{Hz}]$ & ${^2f}_2$ & ${^2f}_{-2}$ & $\delta{f}_{3D}\,[\mathrm{Hz}]$ \\
    \hline
    \hline
    BU0 & 1.446 & 3.956 & 1.596 & 40.0  & 1.565  & 1.565 & 40.0 \\
    \hline
    BU1 & 1.423 & 3.883 & 1.609 & 40.0  & 1.080  & 1.953 & 40.0 \\
    \hline
    BU2 & 1.396 & 3.884 & 1.642 & 40.0  & 0.840   & 2.052 & 40.0 \\
    \hline
    BU3 & 1.360 & 3.915 & 1.680 & 40.0  & 0.640   & 2.123 & 40.0 \\
    \hline
    BU4 & 1.322 & 3.934 & 1.715 & 40.0  & 0.451   & 2.153 & 40.0 \\
    \hline
    BU5 & 1.282 & 3.958 & 1.723 & 40.0  & 0.280   & 2.158 & 40.0 \\
    \hline
    BU6 & 1.258 & 3.998 & 1.744 & 66.67 & 0.130   & 2.143 & 43.5 \\
    \hline
    BU7 & 1.202 & 4.001 & 1.751 & 66.67 & $\times$ & 2.105 & 52.6 \\
    \hline
    \hline  
  \end{tabular}}
\end{table}

\begin{table}
  \caption[Oscillation frequencies (BU sequence, {Cowling} approximation)]{Same as Table~\ref{tab:BUfreqsCFC} but in the Cowling approximation}
  \label{tab:BUfreqsCowling}
  \centering
  {\begin{tabular}{cccccccc}
    \hline
    Model & $F$ & $H_1$ & ${^2f}_0$ & $\delta{f}_{2D}\,[\mathrm{Hz}]$ & ${^2f}_2$ & ${^2f}_{-2}$ & $\delta{f}_{3D}\,[\mathrm{Hz}]$ \\
    \hline
    \hline
    BU0 & 2.683 & 4.561 & 1.881 & 40.0 & 1.881  & 1.881 & 40.0 \\
    \hline
    BU1 & 2.641 & 4.475 & 1.885 & 40.0 & 1.367  & 2.259 & 40.0 \\
    \hline
    BU2 & 2.602 & 4.411 & 1.889 & 40.0 & 1.121  & 2.364 & 40.0 \\
    \hline
    BU3 & 2.561 & 4.403 & 1.883 & 40.0 & 0.918 & 2.443 & 40.0 \\
    \hline
    BU4 & 2.522 & 4.402 & 1.877 & 40.0 & 0.721 & 2.473 & 40.0 \\
    \hline
    BU5 & 2.484 & 4.403 & 1.837 & 40.0 & 0.520 & 2.480 & 40.0 \\
    \hline
    BU6 & 2.461 & 4.401 & 1.762 & 40.0 & 0.348 & 2.477 & 40.0 \\
    \hline
    BU7 & 2.439 & 4.393 & 1.681 & 40.0 & 0.163 & 2.396 & 40.0 \\
    \hline
    \hline  
  \end{tabular}}
\end{table}

\begin{table}
  \caption[Oscillation frequencies from literature (BU sequence, {CFC} and {Cowling} approximation)]{Frequencies extracted from Table II of~\cite{gaertig_oscillations_2008} for axisymmetric modes with the {Cowling}
                                                                                                  approximation, Table 5 of~\cite{dimmelmeier_non-linear_2006} for axisymmetric modes in {CFC},
                                                                                                  digitized with \texttt{engauge-digitizer}~\cite{mitchell_markummitchellengauge-digitizer_2020}
                                                                                                  from Fig. 1 of~\cite{kruger_oscillations_2010} for non-axisymmetric modes with the {Cowling}
                                                                                                  approximation and Table VII of~\cite{kruger_dynamics_2020} for non-axisymmetric modes in full {GR}.}
  \label{tab:BUfreqsliterature}
  \resizebox{14.8cm}{!}{\begin{tabular}{cccccccclll}
    \hline
    Model & \multicolumn{2}{c}{$F$} & \multicolumn{2}{c}{$H_1$} & \multicolumn{2}{c}{${^2f}_0$} & \multicolumn{2}{c}{${^2f}_2$} & \multicolumn{2}{c}{${^2f}_{-2}$} \\
          & {Cowling} & {CFC}  & {Cowling} & {CFC}    & {Cowling} & {CFC}        & {Cowling} & Full {GR}    & {Cowling} & Full {GR}       \\
    \hline
    \hline
    BU0   & 2.679 & 1.458 & 4.561 & 3.971 & 1.890 & 1.586 & 1.881 & 1.578 & 1.881 & 1.578 \\
    \hline
    BU1   & 2.638 & 1.413 & 4.466 & 3.915 & 1.890 & 1.611 & 1.374 & 1.079 & 2.257 & 1.942 \\
    \hline
    BU2   & 2.605 & 1.380 & 4.435 & 3.907 & 1.906 & 1.635 & 1.122 & 0.836 & 2.362 & 2.054 \\
    \hline
    BU3   & 2.570 & 1.343 & 4.409 & 3.921 & 1.895 & 1.669 & 0.909 & 0.636 & 2.428 & 2.118 \\
    \hline
    BU4   & 2.539 & 1.304 & 4.410 & 3.950 & 1.875 & 1.698 & 0.709 & 0.456 & 2.459 & 2.155 \\
    \hline
    BU5   & 2.500 & 1.281 & 4.400 & 3.964 & 1.844 & 1.714 & 0.534 & 0.293 & 2.469 & 2.127 \\
    \hline
    BU6   & 2.484 & 1.219 & 4.392 & 4.010 & 1.794 & 1.729 & 0.355 & 0.146 & 2.467 & 2.182 \\
    \hline
    BU7   & 2.456 & 1.207 & 4.394 & 4.018 & 1.703 & 1.720 & 0.181 & 0.011 & 2.441 & 2.188 \\
    \hline
    \hline  
  \end{tabular}}
\end{table}
\begin{table}
  \caption[Oscillation frequencies: percentage relative difference]{Relative difference in absolute value between the frequencies of Tables~\ref{tab:BUfreqsCFC} and~\ref{tab:BUfreqsCowling} computed with our code
                                                          and the tabulated frequencies reported in Table~\ref{tab:BUfreqsliterature}, in \%.}
  \label{tab:BUfreqsreldiff}
  \resizebox{14.8cm}{!}{\begin{tabular}{cccccccclll}
    \hline
    Model & \multicolumn{2}{c}{$F$} & \multicolumn{2}{c}{$H_1$} & \multicolumn{2}{c}{${^2f}_0$} & \multicolumn{2}{c}{${^2f}_2$} & \multicolumn{2}{c}{${^2f}_{-2}$} \\
          & {Cowling} & {CFC}  & {Cowling} & {CFC}    & {Cowling} & {CFC}        & {Cowling} & Full {GR}    & {Cowling} & Full {GR}       \\
    \hline
    \hline
    BU0   & 0.15 & 0.83 & 0.00 & 0.38 & 0.48 & 0.63 & 0.00 & 0.83     & 0.00 & 0.83 \\
    \hline
    BU1   & 0.11 & 0.70 & 0.20 & 0.82 & 0.27 & 0.12 & 0.51 & 0.09     & 0.09 & 0.56 \\
    \hline
    BU2   & 0.12 & 1.15 & 0.54 & 0.59 & 0.90 & 0.43 & 0.98 & 0.48     & 0.09 & 0.10 \\
    \hline
    BU3   & 0.35 & 1.25 & 0.14 & 0.15 & 0.63 & 0.65 & 0.09 & 0.63     & 0.61 & 0.24 \\
    \hline
    BU4   & 0.67 & 1.36 & 0.20 & 0.40 & 0.64 & 0.99 & 1.66 & 1.11     & 0.57 & 0.09 \\
    \hline
    BU5   & 0.64 & 0.08 & 0.07 & 0.15 & 0.38 & 0.52 & 2.69 & 4.64     & 0.44 & 1.44 \\
    \hline
    BU6   & 0.93 & 3.10 & 0.20 & 0.30 & 1.82 & 0.86 & 2.01 & 12.3     & 0.40 & 1.82 \\
    \hline
    BU7   & 0.70 & 0.42 & 0.02 & 0.57 & 1.31 & 1.77 & 11.0 & $\times$ & 1.88 & 3.94 \\
    \hline
    \hline  
  \end{tabular}}
\end{table}

\section{Summary and conclusion}
\label{sec:ccl}
We presented the main numerical ingredients that compose the new code \texttt{ROXAS}, a general relativistic spectral code dedicated to NS oscillations. Using \texttt{LORENE} as its foundations, we use 
the xCFC formulation of Einstein equations to compute the metric, and the hydrodynamics sector is based upon the previously published formulation that uses only primitive variables, which
is already tried and tested in spherical symmetry. To be able to perform simulations of rotating stars perturbed with non-axisymmetric profiles we implemented and developed a number of techniques which
allowed to perform simulations with spherical coordinates, including the origin in the grid: regularization schemes, a multidomain approach, filters, an adapted mapping that allows to follow
the surface of the star throughout the simulations, two distinct grids that respectively support metric computation and hydrodynamics, along with interpolation schemes, and the use of a well-balanced scheme.
Every tool or technique has been thoroughly tested and therefore allowed to perform simulations of polytropic stars and stars described with an analytical representation of a realistic
EoS. The standard test beds for NS general relativistic evolution codes such as the migration test and the collapse to a black hole were previously presented in~\cite{servignat_new_2023},
and we here presented the frequency extraction for axisymmetric and non-axisymmetric simulations. The results were compared to previously published results and were found to be very accurate.
In particular, we emphasize that although the simulations are performed under the CFC approximation, the reference frequencies for non-axisymmetric simulations were computed in full General
Relativity, and the frequency extraction is very satisfactory. This means that the CFC approximation seems to be very good for frequency extraction; using it represents a significant gain of computational resources compared to a possible
implementation of General Relativity. The computational cost of the current version of the code is dominated by the interpolation, therefore efforts will be directed to the improvement of
interpolation schemes. All in all, \texttt{ROXAS} fulfills its goal to be a lightweight code that runs on office computers and provides the possibility to perform parametric studies of NS oscillations
using a wide panel of EoS. Future versions will also support EoS with more than one argument by taking out-of-$\beta$-equilibrium and temperature effects, relax the spatial symmetries, go
beyond CFC and support differential rotation profiles, for example those of~\cite{iosif_equilibrium_2021} that are relevant for simulations of a post-coalescence object like a HMNS.

\ack
The authors gratefully acknowledge the Italian Instituto Nazionale de Fisica Nucleare (INFN), the French Centre National de la Recherche Scientifique (CNRS) and the Netherlands Organization for Scientific Research for the construction and operation of the Virgo detector and the creation and support of the EGO consortium. 
GS thanks all the ROC team of LUTH for creating a work environment conducive to the success of his PhD.

\appendix
\section{Derivation of the impermeable boundary condition}
\label{app:derivationBC}
In this appendix we show the derivation of the expression of the coordinates of a normal vector defined on an arbitrarily deformed sphere.
We then apply this to write the impermeable BC with a given coordinate velocity field $(v^r,\,v^\theta,\,v^\phi)$.
To find out how to write the condition, we first assume that the deformed boundary is mapped from $\mathbf{S}^2$, the unit sphere of $\R^3$ by:
\begin{equation}
  \mathbf{R} : \mathbf{x}\in\mathbb{S}^2\mapsto R(\theta,\phi)\mathbf{x}.
\end{equation}
Then we compute the (rescaled) angular gradient of the radius:
\begin{equation}
  \mathbf{g} = \frac{\bnabla_{\theta\phi}R}{R},
\end{equation}
where
\begin{equation}
  \bnabla_{\theta\phi} = \left(\begin{array}{c}
  0 \\ \drond[]{\theta} \\ \frac{1}{\sin\theta}\drond[]{\phi}
  \end{array}\right)
\end{equation}
Then we compute the tangent vector:
\begin{equation}
  \mathbf{h} = \mathbf{g} - (\mathbf{g}\cdot\mathbf{x})\mathbf{x}.
\end{equation}
Finally, the (unnormalised) normal vector to the deformed sphere at $\mathbf{R}(\mathbf{x})$ is:
\begin{equation}
  \mathbf{n} = \mathbf{x}-\mathbf{h}.
\end{equation}
Its coordinates in the orthonormal spherical triad are:
\begin{eqnarray}
  n_r & = x_r - h_r = 1-(g_r-g_r) = 1, \\
  n_\theta & = x_\theta-h_\theta = -g_\theta = -\frac{1}{R}\drond[R]{\theta}, \\
  n_\phi & = x_\phi - h_\phi = -g_\phi = -\frac{1}{R\sin\theta}\drond[R]{\phi}.
\end{eqnarray}
Therefore, with $\mathbf{v}_\mathrm{border} = \drond[R]{t}\mathbf{\hat{e}_r}$, the impermeable {BC} is expressed as:
\begin{equation}
  \label{eq:impermeableBC_appendix}
  v_r(R(t),t) - \frac{v_\theta(R(t),t)}{R}\drond[R]{\theta} - \frac{v_\phi(R(t),t)}{R\sin\theta}\drond[R]{\phi} = \drond[R]{t}.
\end{equation}
This expression was found by~\cite{uryu_existence_1996} when looking for non-axisymmetric equilibrium configurations of self-gravitating objects.
We challenged the expression numerically, the results can be found in~\ref{app:numtestBC}.

\section{Numerical test of the impermeable boundary condition}
\label{app:numtestBC}

In this appendix we provide numerical evidence that Equation~(\ref{eq:impermeableBC_appendix}) corresponds to the correct impermeable BC for a general velocity field
inside a self-gravitating object. To do so, we perform the simulation of a rotating ellipsoid and show a comparison with the theoretical rotating configuration,
which can be derived analytically in some simple cases. We start with some definitions, write Equation~(\ref{eq:impermeableBC_appendix}) for a rotating ellipsoid around a given
axis, derive the analytical formula with which we compare the code, and show some numerical results.

\subsection{Introduction}
\subsubsection{Definition}
An ellipsoid is a surface of $\R^3$ that can be obtained with an affine transformation applied on a sphere. Its general definition in Cartesian coordinates is:
\begin{equation}
  \label{eq:defellipsoid}
  \frac{x^2}{a^2} + \frac{y^2}{b^2} + \frac{z^2}{c^2} = 1,
\end{equation}
where $a,\,b,\,c$ are the length of the three principal axes.

\subsubsection{Parametrization}
Using polar spherical coordinates $(r,\,\theta,\,\phi)$, the surface of the ellipsoid can be parametrized as:
\begin{equation}
  \label{eq:param1ellipsoid}
  R(\theta,\,\phi) = \left(\frac{\cos^2\phi\sin^2\theta}{a^2} + \frac{\sin^2\phi\sin^2\theta}{b^2} + \frac{\cos^2\theta}{c^2}\right)^{-1/2},
\end{equation}
or, alternatively:
\begin{equation}
  \label{eq:param2ellipsoid}
	R(\theta,\,\phi) = \frac{abc}{\sqrt{c^2\sin^2\theta(b^2\cos^2\phi + a^2\sin^2\phi) + a^2b^2\cos^2\theta}}.
\end{equation}

\begin{figure*}
  \begin{minipage}[t]{.95\textwidth}
    \begin{minipage}[b]{.51\textwidth}
      \centering
      \includegraphics[width=\columnwidth]{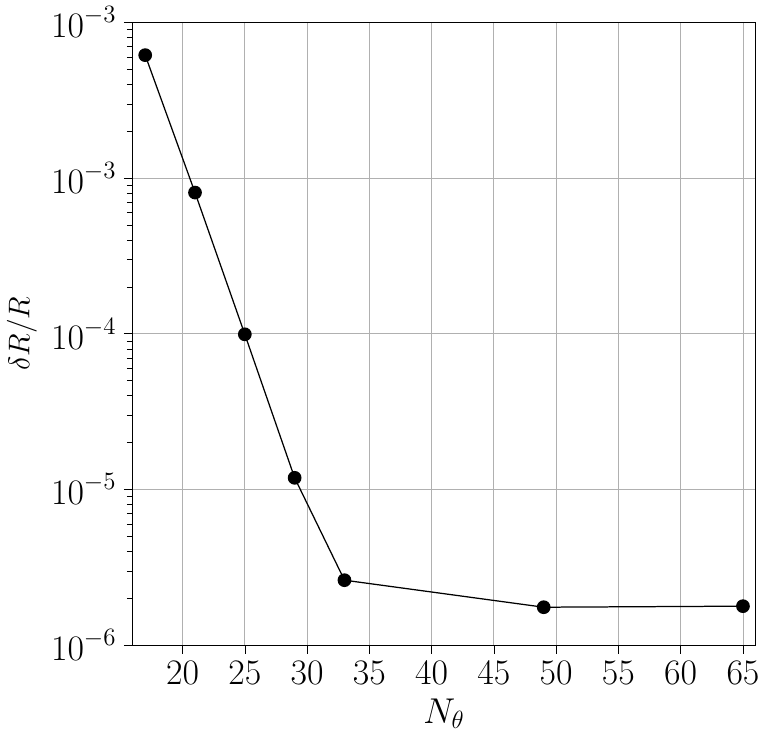}
    \end{minipage}
    \begin{minipage}[b]{.48\textwidth}
      \centering
      \includegraphics[width=\columnwidth]{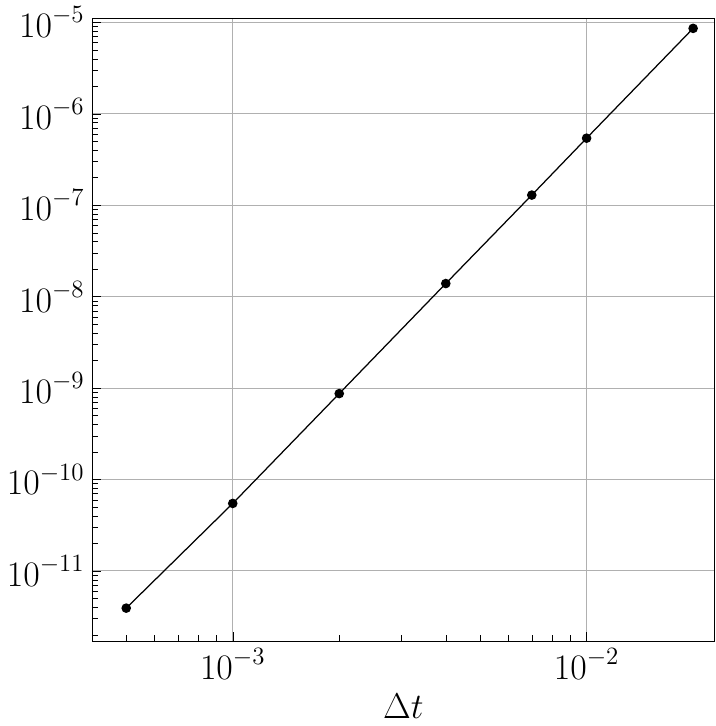} 
    \end{minipage}
  \end{minipage}
  \caption[Convergence plots]{Convergence plots for a rotating ellipsoid. The quantity plotted is defined by Equation~(\ref{eq:reldiffellipsoid}). On the left, we fixed $t_\mathrm{max}=100$, $\Delta{t}=10^{-3}$ and $\Np=32$ and varied $\Nt$. On the right, we fixed $t_\mathrm{max}=10$, $\Nt=65$ and $\Np=64$
  and varied $\Delta{t}$.}
  \label{fig:convergenceplotrotatingellipsoid}
\end{figure*}

\subsection{Rotation field}
Now that we have given the parametrization of an ellipsoid, let us turn to the velocity field that describes a rotation around a given axis. We define it as
\begin{equation}
  \mathbf{v} = R(\theta,\,\phi)\bOmega,
\end{equation}
where $\bOmega$ is the rotation vector, of which the norm $\Omega=|\bOmega|$ is the angular rotation rate and direction is defined with respect to some point of the ellipsoid
$\mathbf{M}=R\mathbf{\hat{e}_r}$ and the chosen rotation axis. It means that $\mathbf{v}=\Omega\mathbf{M}\times\mathbf{\hat{e}}$, where $\mathbf{\hat{e}}$ is a generic axis in $\R^3$.
This yields three possible rotation laws associated to the three Cartesian axes (of which the coordinates are given in the spherical orthonormal basis):
\begin{eqnarray}
	\bOmega_x/\Omega  & = \mathbf{\hat{e}_x}\times\mathbf{\hat{e}_r} = \left(\begin{array}{c}
		0 \\ -\sin\phi \\ -\cos\theta\cos\phi
	\end{array}\right), \\
	\bOmega_y/\Omega  & = \mathbf{\hat{e}_y}\times\mathbf{\hat{e}_r} = \left(\begin{array}{c}
		0 \\ \cos\phi \\ -\cos\theta\sin\phi
	\end{array}\right), \\
	\bOmega_z/\Omega  & = \mathbf{\hat{e}_z}\times\mathbf{\hat{e}_r} = \left(\begin{array}{c}
		0 \\ 0\\ \sin\theta
	\end{array}\right),
\end{eqnarray}
and therefore three PDEs that describe the evolution of a rotating ellipsoid:
\begin{eqnarray}
  \frac{1}{\Omega}\drond[R]{t} = & \sin\phi\drond[R]{\theta} + \frac{\cos\phi}{\tan\theta}\drond[R]{\phi} & \mbox{ (rotation around $x$ axis)}, \\
  \frac{1}{\Omega}\drond[R]{t} = & - \cos\phi\drond[R]{\theta} + \frac{\sin\phi}{\tan\theta}\drond[R]{\phi} & \mbox{ (rotation around $y$ axis)}, \\
  \frac{1}{\Omega}\drond[R]{t} = & - \drond[R]{\phi} & \mbox{ (rotation around $z$ axis)}. \\
\end{eqnarray}
For a given constant angular rotation rate $\Omega$, the quantity $\alpha$, defined as the angle between the ellipsoid and a given axis and assumed to be initially zero, varies as
\begin{equation}
  \alpha(t) = \Omega t. \label{eq:rotationlawy}
\end{equation}
With this convention, the ellipsoid rotates counterclockwise with respect to the chosen axis.

\subsection{Numerical results}

For all simulations we set $\Omega=1$. The angular derivatives are computed thanks to the spectral
representation of $R(\theta,\,\phi)$, and the evolution is performed thanks to an RK4 scheme. Rotation around the $z$ axis is
straightforward, however when the ellipsoid rotates around the $x$ or the $y$ axis, it crosses the $z$ axis on which the angular coordinates are singular.
To cure this issue, we perform the regularization process described in Sec~\ref{subsubsec:regularizationaxis}.
We compute the rotation around the $y$ axis for an ellipsoid with the parameters $a=2$, $b=1.5$, $c=1$. On Figure~\ref{fig:convergenceplotrotatingellipsoid}
we show the results of a convergence test. We measure the error on a simulation by computing
\begin{equation}
  \label{eq:reldiffellipsoid}
  \delta{R}/R = \max\limits_{t\in[0,t_\mathrm{max}],\,\theta\in[0,\pi],\,\phi\in[0,2\pi)}\left|\frac{R_\mathrm{numerical}(t,\,\theta,\,\phi) - R_\mathrm{analytical}(t,\,\theta,\,\phi)}{R_\mathrm{analytical}(t,\,\theta,\,\phi)}\right|,
\end{equation}
where $R_\mathrm{numerical}$ is the ellipsoid configuration computed with the code and $R_\mathrm{analytical}$ the one computed by writing the general expression of a tilted ellipsoid around a given
axis, and assuming the rotation law~(\ref{eq:rotationlawy}).

By fixing $\Delta{t}$ and $\Np$ to vary $\Nt$ we show the exponential convergence of the spatial scheme,
which is here shown by the presence of a straight line in the semilog plot before reaching the plateau at $\Nt=33$, and then
fixing $\Nt$ and $\Np$ to vary $\Delta{t}$ we show the order of the time scheme: since we use the RK4 scheme we expect a $\mathcal{O}(\Delta{t}^4)$ behaviour; the log-log plot
shows a straight line with a slope of 4, which validates the convergence test for the time integration scheme.

\section{Scalar field matching}
\label{app:matchwires}
In this appendix we describe the so-called tau method to match a scalar field $f$ between a nucleus and a single shell by imposing the continuity of $f$ and $\partial_rf$. 
The spectral expansion of $f$ is:
\begin{equation}
	f_\mathrm{nuc}(\xi) = \sum\limits_{k=0}^{N_{r,\mathrm{nuc}}-1} f_{\mathrm{nuc},k}T_{2k}(\xi)
\end{equation}
in the nucleus, and:
\begin{equation}
	f_\mathrm{sh}(\xi) = \sum\limits_{k=0}^{N_{r,\mathrm{sh}}-1} f_{\mathrm{sh},k}T_k(\xi)
\end{equation}
in the shell.
Writing the continuity of the field and of its first derivative allows writing a linear system with $f_{n,N_{r,\mathrm{nuc}}-1}$ and $f_{s,N_{r,\mathrm{sh}}-1}$ as unknowns by writing the following conditions:
\begin{eqnarray}
	f_\mathrm{nuc}(1) & = f_\mathrm{sh}(-1), \label{eq:functioncontinuity}\\
	\frac{1}{\alpha_\mathrm{nuc}}f_\mathrm{nuc}'(1) & = \frac{1}{\alpha_\mathrm{sh}}f_\mathrm{sh}'(-1),\label{eq:derivativecontinuity}
\end{eqnarray}
and making use of the following properties of the Chebyshev polynomials~\cite{hesthaven_spectral_2007}:
\begin{equation}
	\forall k\in\N,\, T_k(\pm1) = (\pm1)^k,\, T_k'(\pm1) = (\pm1)^{k+1}k^2.
\end{equation}
which allow to rewrite the junction conditions:
\begin{eqnarray}
	\sum\limits_{k=0}^{N_{r,\mathrm{nuc}}-1} f_{\mathrm{nuc},k} & = \sum\limits_{k=0}^{N_{r,\mathrm{sh}}-1} (-1)^kf_{\mathrm{sh},k}, \\
	\frac{4}{\alpha_\mathrm{nuc}}\sum\limits_{k=0}^{N_{r,\mathrm{nuc}}-1} k^2f_{\mathrm{nuc},k} & = \frac{1}{\alpha_\mathrm{sh}}\sum\limits_{k=0}^{N_{r,\mathrm{sh}}-1} (-1)^{k+1}k^2f_{\mathrm{sh},k}.
\end{eqnarray}
We can rewrite this in a matrix form $AX=B$ with:
\begin{eqnarray}
	A & =\left(\begin{array}{cc}
		1 & -1 \\
		\frac{4(N_{r,\mathrm{nuc}}-1)^2}{\alpha_\mathrm{nuc}} & \frac{(N_{r,\mathrm{sh}}-1)^2}{\alpha_\mathrm{sh}}
	\end{array}\right),\\	
	X & =\left(\begin{array}{c}
		f_{n,N_{r,\mathrm{nuc}}-1} \\ f_{s,N_{r,\mathrm{sh}}-1}
	\end{array}\right),\\
	 B & =\left(\begin{array}{c}
		\sum\limits_{k=0}^{N_{r,\mathrm{nuc}}-2} f_{\mathrm{nuc},k} - \sum\limits_{k=0}^{N_{r,\mathrm{sh}}-2} (-1)^kf_{\mathrm{sh},k} \\ \frac{4}{\alpha_\mathrm{nuc}}\sum\limits_{k=0}^{N_{r,\mathrm{nuc}}-2} k^2f_{\mathrm{nuc},k} - \frac{1}{\alpha_\mathrm{sh}}\sum\limits_{k=0}^{N_{r,\mathrm{sh}}-2} (-1)^{k+1}k^2f_{\mathrm{sh},k}
	\end{array}\right),
\end{eqnarray}
then solve for $X$ by computing the inverse of $A$, and finally replace the last coefficients of $f_\mathrm{nuc}$ and $f_\mathrm{sh}$ by the first and second coefficients of the vector $X$. We make use of the explicit inverse of a 2x2 matrix to write:
\begin{equation}
	A^{-1} = \left(\begin{array}{cc}
		\frac{\alpha_\mathrm{nuc}(N_{r,\mathrm{sh}}-1)^2}{\alpha_\mathrm{nuc}(N_{r,\mathrm{sh}}-1)^2 + 4\alpha_\mathrm{sh}(N_{r,\mathrm{nuc}}-1)^2} & -\frac{\alpha_\mathrm{nuc}\alpha_\mathrm{sh}}{\alpha_\mathrm{nuc}(N_{r,\mathrm{sh}}-1)^2 + 4\alpha_\mathrm{sh}(N_{r,\mathrm{nuc}}-1)^2} \\
		4\frac{\alpha_\mathrm{sh}(N_{r,\mathrm{nuc}}-1)^2}{\alpha_\mathrm{nuc}(N_{r,\mathrm{sh}}-1)^2 + 4\alpha_\mathrm{sh}(N_{r,\mathrm{nuc}}-1)^2} & -\frac{\alpha_\mathrm{nuc}\alpha_\mathrm{sh}}{\alpha_\mathrm{nuc}(N_{r,\mathrm{sh}}-1)^2 + 4\alpha_\mathrm{sh}(N_{r,\mathrm{nuc}}-1)^2}
	\end{array}\right).
\end{equation}

\section*{References}
\bibliographystyle{unsrt}
\bibliography{biblio}

\end{document}